\documentclass[10pt, conference, letterpaper]{IEEEtran}
\IEEEoverridecommandlockouts

\usepackage{cite}
\usepackage{amsmath,amssymb,amsfonts}
\usepackage{algorithmic}
\usepackage{graphicx}
\usepackage{textcomp}
\usepackage{xcolor}
\usepackage{caption}
\usepackage{hyperref}
\usepackage{cleveref}
\usepackage{verbatim}
\usepackage{orcidlink}

\def\BibTeX{{\rm B\kern-.05em{\sc i\kern-.025em b}\kern-.08em
    T\kern-.1667em\lower.7ex\hbox{E}\kern-.125emX}}
\newcommand{\GlobalpingSystemName}{Globalping}
\newcommand{\CoreAPIServer}{Core API Server}
\newcommand{\GeoIPService}{GeoIP Service}
\newcommand{\ProbeManagementModule}{Probe Management Module}
\newcommand{\FrontEnd}{front end}
\newcommand{\RQOne}{RQ1: How does \GlobalpingSystemName{} compare to similar tools?}
\newcommand{\RQTwo}{RQ2: How is users' adoption of \GlobalpingSystemName{}?}
\newcommand{\RQThree}{RQ3: How usable is \GlobalpingSystemName{} compared to RIPE Atlas?}
\begin{document}

\title{\GlobalpingSystemName: A Community-Driven, Open-Source Platform for Scalable, Real-Time Network Measurements\\
}

\author{\IEEEauthorblockN{1\textsuperscript{st} Berkay Kaplan \orcidlink{0000-0002-4365-7606}}
\IEEEauthorblockA{\textit{Rutgers Business School – Newark and New Brunswick} \\
\textit{Rutgers University}\\
Piscataway, New Jersey, USA \\
berkay.kaplan@rutgers.edu}
}
\maketitle

\begin{abstract}
We present \GlobalpingSystemName{}, an open-source, community-driven platform designed to support architecturally scalable, real-time global network measurements. It democratizes access to network diagnostics by enabling every user, including both non-technical and technical users, as well as companies, to perform ping, traceroute, and DNS lookups via a globally distributed network of user-hosted probes, either through the intuitive \GlobalpingSystemName{} front-end or the REST API. Unlike RIPE Atlas and similar platforms, official integrations with other platforms, such as Slack and GitHub, support real-time monitoring and collaborative workflows. \GlobalpingSystemName{} introduces a combination of features that, to our knowledge, are not jointly supported by existing global network diagnostics platforms, including official support for containerized probes. We conducted a user study with 40 participants. We observed a System Usability Scale (SUS) score much higher than RIPE ATLAS for a single and simple diagnostic ping task, chosen as a representative baseline operation common to all global network diagnostic platforms.
\end{abstract}

\begin{IEEEkeywords}
Network Monitoring, Distributed Systems, Network Diagnostics, Remote Probes, Ping Testing, Latency Measurement, Network Troubleshooting, Geographically Distributed Nodes
\end{IEEEkeywords}

\section{Introduction}

With the expansion of the global Internet infrastructure in recent years, real-time network diagnostics (ND) and monitoring have become increasingly important for both technical and non-technical users, as networks now sit at the center of almost everything people and organizations do \cite{b8}. Our heavy dependence on the internet and cloud services, the growth of remote work and online learning, and the increasing complexity of networks make real-time ND critical. Fortunately, we have tools to assist with this task.

Although these tools assist with ND, most serve only a select few professionals rather than broader groups, as they have a high barrier to entry. They also don't allow real-time measurements. For instance, RIPE Atlas measurements are near-real-time, which are fast enough to reflect current conditions, but they are not true real-time streaming data like you'd get from an active network monitor inside your own infrastructure. In this paper, we define real-time measurements as measurements whose results are delivered continuously with sub-second latency as they are produced. Near-real-time measurements are returned with short but discrete delays due to scheduling, batching, or polling, and these delays can be particularly significant when you require immediate awareness, rapid response, or when a problem lasts only a few seconds and appears and disappears quickly. There is also no major platform designed to be natively embedded into modern CI/CD pipelines, such as GitHub Actions, or communication tools like Slack and Discord. This embedding would move network measurement from a standalone research task into an integrated part of the software development lifecycle.

This paper introduces \GlobalpingSystemName{}, an open-source community-driven platform that provides democratic access to architecturally scalable real-time network measurements. Unlike similar tools, \GlobalpingSystemName{} empowers all users, including experts and non-technical users, to perform networking diagnostics such as ping, traceroute, and DNS lookups via a globally distributed network of user-hosted, containerized probes.

\GlobalpingSystemName{} illustrates a flexible, community-powered architecture, making it a valuable resource for real-time Internet and network performance monitoring and network troubleshooting. \GlobalpingSystemName{} makes the following contribution to the literature.
\begin{itemize}
  \item A platform democratizing ND through low-barrier access, designed to be natively embedded into modern CI/CD pipelines and communication tools, embracing containerization and ephemeral infrastructure instead of using stable, long-term hardware probes, enabling real-time measurements with a WebSocket architecture.
  \item A comparative study of \GlobalpingSystemName's advantages over similar tools.
  \item A user study comparing \GlobalpingSystemName{} with RIPE Atlas.
\end{itemize}

In addition to being the first to introduce several features, such as real-time measurements, our user study reveals that Globalping is superior in usability and accessibility compared to traditional global ND tools. The study involved 40 users with completely random backgrounds and found that Globalping had a SUS score of 65.62, significantly higher than RIPE Atlas's score of 42.38. Users performed a ping task in half the time in Globalping, which is 3.23 minutes on average, compared to RIPE Atlas, which took 6.87 minutes, highlighting its usefulness in accelerating and democratizing access to real-time ND.

In addition to our results, the following section provides background on some concepts. Section 3 will focus on \GlobalpingSystemName's high-level capabilities, node setup, and official integrations that can expedite workflows. Section 4 will elaborate on the implementation of \GlobalpingSystemName{}, its architecture, and security practices. Section 5 presents the results of our comparative and user study on \GlobalpingSystemName{}. We will interpret those results in the discussion section, present academic papers related to our work, and finally conclude our work.

\section{Background}

One of the essential network debugging tools is PING, which uses two ICMP query messages: ICMP (ECHO request) and ICMP (ECHO reply) \cite{b13}. When a source makes an ICMP (ECHO (PING) request) to another host, according to the RFC 0792 guidelines, that host must respond with an ICMP (ECHO (PING) reply) after receiving the request from that source \cite{b13}.

Another vital network debugging tool is Traceroute, a networking operation that returns an IP address and a hostname for each network layer device, such as a router, along the path from a source to a destination in a network \cite{b12}. Each router examines the Time-to-Live (TTL) of an IP packet it receives \cite{b12}. If the TTL is greater than 1, the router decrements it and forwards the packet to the next router on the path to its destination \cite{b12}. If the TTL is equal to one, the router drops the packet and sends an ICMP Time Exceeded message back to the packet source \cite{b12}. Traceroute sends several packets with incremental TTLs, starting at 1, and receives ICMP Time Exceeded messages from each router along the path to the destination. These messages allow the traceroute tool to infer the IP address of each router on the path to the destination, thus building a network map \cite{b12}.

In addition to networking concepts, we also conducted an SUS study, a specific type of user study that uses the System Usability Scale (SUS). It is a short and reliable questionnaire for assessing perceived system or product usability. Researchers administer the 10-item SUS questionnaire after user interaction with a system, gather quantitative user satisfaction data, and transform it into a single overall ease-of-use measure, which emerges as a score on a 0-100 scale. A higher score suggests higher usability and helps identify areas for improvement, making the case for investing in the subject UX.

\section{Related Work}

MacMillan et al. focus on the application of network performance measurement metrics, such as download and upload speeds, and elaborate on M-Lab's NDT7 \cite{b18}. It provides feedback on network throughput and is the most widely used tool in the repertoires of consumers, regulators, and ISPs for measuring network quality \cite{b18}. A side-by-side comparison between Ookla's Speedtest and NDT7 revealed that, though the two tools report equal speeds for most setups, their perceived throughput varies under networks with hefty latency. In contrast, Ookla reports consistently higher speeds \cite{b18}. This paper focuses mainly on Internet speed tests, while \GlobalpingSystemName{} also covers other use cases, such as mapping network paths to a server from the user's chosen location.

DiagNet is a topology-agnostic machine learning-based diagnosis of problems in Internet-scale services that does not require any predefined lists of possible causes at training time \cite{b19}. This paper applies image processing techniques to network and system metrics for root-cause analysis in large network environments \cite{b19}. It has a recall of up to 73.9\% when diagnosing problems during experiments, including causes only seen at inference time \cite{b19}. Although machine learning can improve ND productivity, giving users manual control can yield higher accuracy. However, machine learning can provide preliminary results for engineers, so they can start ahead of point zero when troubleshooting.

Yang et al. compiled a survey of recent methods for network fault diagnosis, highlighting advances in network fault detection and diagnosis \cite{b20}. Several approaches are reviewed, including knowledge-based approaches, and the contribution of accurate information collection and analysis in network fault diagnosis is underlined \cite{b20}. The researchers also describe some open issues and directions for research and development in this area \cite{b20}. We did not find popular tools such as RIPE Atlas in this paper, although it was released 7 years before the publication of this survey.

Although there are some helpful tools like RIPE Atlas, NDT7, DiagNet, and NetGraf available today that offer network diagnosis functionality, they usually come with caveats: requiring physical hardware deployment, specialized network conditions, or complex configuration. Globalping aims to ease these pains by implementing an open-source, open-access system based on community contributions to perform ND worldwide without specialized hardware or complex configurations.

\section{System Functions and API Overview}

\begin{figure*}[ht]
\centering
\includegraphics[width=0.6\textwidth]{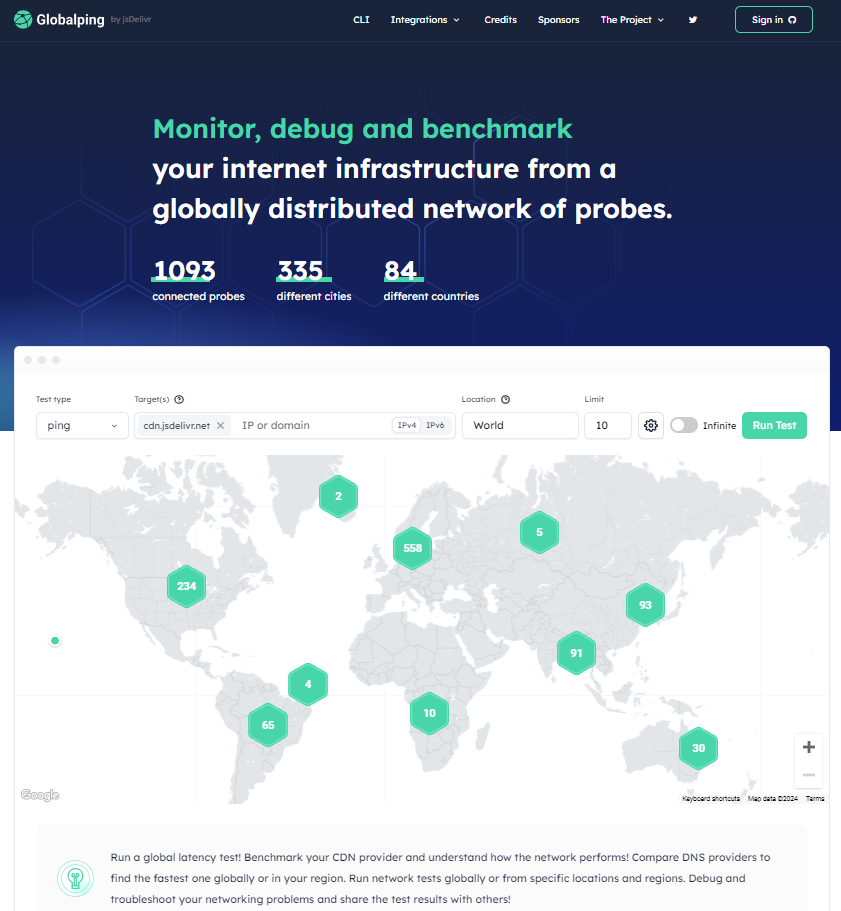}
\caption{\GlobalpingSystemName{}'s user interface (\FrontEnd{}) dashboard.}
\label{figure:front-end}
\end{figure*}

Users can use \GlobalpingSystemName{} through its user interface (\FrontEnd{}) at \url{https://globalping.io/}, REST API endpoints at \url{https://api.globalping.io/v1}, or the CLI. \GlobalpingSystemName{}'s \FrontEnd{} focuses on simplicity by allowing anyone to run measurements without login by clicking the "Run Test" button in Figure \ref{figure:front-end} after setting the parameters.

\GlobalpingSystemName{}'s CLI is a command-line tool to interact with \GlobalpingSystemName{}'s REST API endpoints to carry out network measurements and authenticate using either a browser-based flow or a token. The CLI is helpful for scripting and maintaining a command history.

\GlobalpingSystemName{}'s REST API uses HTTPS to execute network commands, such as ping or traceroute, programmatically, and to retrieve data to embed \GlobalpingSystemName{} functionality into user tools or dashboards. While the user must poll the HTTP endpoint of \GlobalpingSystemName{} to retrieve their measurement results, \GlobalpingSystemName{} uses WebSockets internally, enabling real-time interaction for continuous monitoring with instantaneous data presentation and measurement results at minimum latency.

\GlobalpingSystemName{} is community-driven and uses users from around the world to power its globally distributed probe infrastructure. Thus, it built an economy that rewards users with credits for hosting probes and allows them to spend those credits to conduct measurement tests on other probes. Anonymous users who don't host probes or have an account can still run measurements with a limit of 250 free tests per hour and 50 probes per measurement. When users run a measurement, they provide the number of results they want. For example, running ping measurements from 10 locations in Germany means users will request 10 different probes, which will cost 10 credits. Users with an account will have 500 free tests per hour and 500 probes per measurement. Users only need a GitHub account to increase their limit. Users can also earn 150 tests for each probe they host that remains online all day and receive more than 2000 tests per \$1 donated.

\subsection{Measurement Capabilities}

Users have to enter two parameters to run most measurements: the target IP or URL and the origin location, such as Germany. Each measurement type has optional configurations that users can set programmatically or by \FrontEnd{} with the gear icon in Figure \ref{figure:front-end}. For example, the user can decide how many packets they want for their ping measurement. Without any configurations, \GlobalpingSystemName{}'s most simple capabilities are:

\begin{itemize}
\item Ping measurements output the RTT for a set of packets sent to and from a target IP address or domain. It outputs statistics, including packet loss and minimum, average, and maximum RTT.

\item Traceroute measurements visualize a packet's path to a target by showing each hop (router) along the way, along with the time required to reach each hop.

\item HTTP measurements check the availability of a web service by sending GET or HEAD requests. It also measures the time required for DNS resolution and for establishing a TCP connection to a web service.

\item Multi-hop measurement is similar to traceroute, showing the path packets take. In addition, they can measure other protocols, such as HTTP or TCP.

\item DNS measurements are similar to the domain information groper (dig) command. These measurements can verify and troubleshoot DNS problems and perform DNS lookups.

\end{itemize}

\subsection{Node Setup}

\GlobalpingSystemName{} has a distributed architecture through a network of community-hosted nodes. Geographic diversity is vital for \GlobalpingSystemName{} to enable users to conduct network performance measurements from a variety of locations worldwide, but the distribution of nodes can be biased toward densely populated areas, such as urban areas. To mitigate this bias, \GlobalpingSystemName{} allows users to select specific countries, cities, or regions when making measurement requests, and then shows the probe location for each result so users can evaluate geographic representation and potential bias.

Users need a server or virtual machine with at least one vCPU and 512 MB of RAM. \GlobalpingSystemName{} only requires the user to install Docker and pull its image. If users need to set up multiple nodes, \GlobalpingSystemName{} provides an automated script to install probes on Linux servers. \GlobalpingSystemName{} also has a Podman alternative for users who do not use Docker.

To the best of our knowledge, Globalping is the first platform of its kind to officially adopt containerization, although other tools, such as RIPE ATLAS, have unofficial and community support for containerization. This approach significantly lowers the barrier to entry for hosting a probe. If users need to set up multiple nodes, Globalping provides an automation script to install probes on Linux servers.

\subsection{Official Integrations}

\GlobalpingSystemName{} includes an official Slack integration that lets teams run measurements and receive real-time updates, results, and probe statuses without leaving their Slack channels. Such Slack integration enables collaborative troubleshooting and the immediate sharing of results, helping teams take quick action against network issues. There is also an official integration with the GitHub bot that can interact with the \GlobalpingSystemName{} network in GitHub comments. \GlobalpingSystemName{} developers are currently working on developing more integrations, including ChatGPT, Zapier, GitHub Actions, IFTTT, and NetBox, turning \GlobalpingSystemName{} from a standalone ND tool into a practical, scalable, and widely usable Internet measurement service for many different users and workflows

\section{Implementation}

\begin{figure}[ht!]
    \centering
    \includegraphics[width=.4\textwidth]{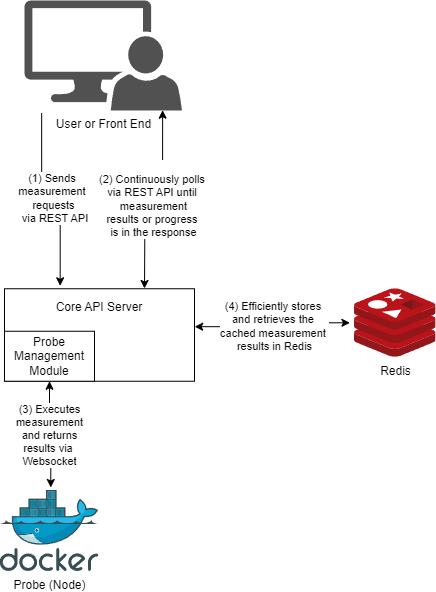}\hfill
    \caption{\GlobalpingSystemName{}'s workflow for executing measurements with or without the \FrontEnd{}.}
    \label{figure:workflow}
\end{figure}

\GlobalpingSystemName{}'s main services are the \CoreAPIServer{} and \ProbeManagementModule{}. The \CoreAPIServer{} follows a monolithic architecture with the  \ProbeManagementModule{}, meaning they're a single service. Globalping also utilizes \FrontEnd{}, \GeoIPService{}, and Redis

\subsection{System Architecture Overview}

The \CoreAPIServer{} provides REST API endpoints for users to submit measurement requests. It maintains the measurement settings and communicates with other services as the central server. The \CoreAPIServer{} uses the \GeoIPService{} to determine the location of IP addresses from geolocation providers like MaxMind, IPInfo, and Fastly. The \CoreAPIServer{} also uses Redis for various tasks, such as storing measurement results. 

In addition to REST API requests, \CoreAPIServer{} can also receive measurement requests from \FrontEnd{}, a web interface that allows users to configure their measurements and view results. The \FrontEnd{} interacts with the \CoreAPIServer{}'s REST API endpoints to visualize the measurement data. To support high request volumes, the \CoreAPIServer{} is horizontally scalable and can run multiple stateless instances behind a load balancer.

The \FrontEnd{} initiates the measurements by making REST API requests to the \CoreAPIServer{}. The \CoreAPIServer{} triggers the \ProbeManagementModule{}. \GlobalpingSystemName{}'s \ProbeManagementModule{} manages a set of preconfigured and geographically distributed probes to execute the measurements. The \ProbeManagementModule{} connects to a probe via WebSocket and sends the measurement job details. The probe executes the job and returns the result to \ProbeManagementModule{} via the WebSocket connection. The \ProbeManagementModule{} delivers the results to the \CoreAPIServer{}, which returns them to the \FrontEnd{}.

The \FrontEnd{} receives real-time updates for user measurement requests from the \CoreAPIServer{} by polling its REST API endpoints every second. This polling technique is useful for displaying long-running measurements, such as traceroute or multihop, in real-time. The \CoreAPIServer{} uses Redis to cache and quickly handle the measurement results. The \FrontEnd{} also uses Redis to manage log-in, sessions, and permissions with its separate API. \GlobalpingSystemName{}'s complete workflow for measurement requests is shown in Figure \ref{figure:workflow}. If the user does not use the \FrontEnd{} and make measurement requests directly to the \CoreAPIServer{}'s REST API, the user would have to continuously poll the \CoreAPIServer{} to retrieve their results.

The \CoreAPIServer{} also uses Redis to cache and store real-time updates for network measurement results, probe statuses, and geolocation information. Redis contains the latest measurement status updates, but only \CoreAPIServer{} can store or retrieve data from it. The results have a time-to-live (TTL) set in Redis, which ensures that \GlobalpingSystemName{} can handle multiple concurrent measurements. Redis is useful for fast lookups of frequently accessed data and keeping the state in sync across \GlobalpingSystemName{}'s distributed components. 

\subsection{Node Infrastructure and Management}

Each node corresponds to a probe that executes the measurement request it receives from \ProbeManagementModule{} within \CoreAPIServer{}. \GlobalpingSystemName{} deploys the probes with Docker containers, making them platform-agnostic and adaptable to cloud environments such as AWS. Whenever a new probe becomes online, it sends its IP address and unique node identifier to \CoreAPIServer{}. The \CoreAPIServer{} will create a new record for it in its node registry and resolve its geographic information, including the probe's city and country, using the \GeoIPService{}. Once a node has registered, it must send a regular heartbeat to \ProbeManagementModule{} to indicate that it is online. Otherwise, \ProbeManagementModule{} will mark it as offline.

If a developer wants to contribute to and debug \GlobalpingSystemName{} nodes, they'd need to clone the repo and set the probe's Docker container environment variables. The critical environment variables are API\_HOST, ADMIN\_KEY, and SYSTEM\_API\_KEY. These configurations tell the node which \CoreAPIServer{}'s \ProbeManagementModule{} it needs to connect to and provide the credentials for authentication, which returns a session token to the node to avoid repeatedly sending its API key. The node initiates a WebSocket connection, which is the primary channel for all interactions, to the \ProbeManagementModule{}. The setup instructions in the \GlobalpingSystemName{} documentation on the user interface and in the GitHub README explain how to generate API keys, set environment variables, and properly configure the nodes.

\subsection{Measurement Scheduling and Assignment}

\subsubsection{Measurement Request Handling}
Users can trigger a measurement via \CoreAPIServer{}'s REST API endpoints, the CLI, or \FrontEnd{}. All requests must include a set of parameters, such as the measurement type (e.g., ping, traceroute), the origin city, the target URL, and any request-specific options. \GlobalpingSystemName{} will validate each request and, if invalid, return a 400 Bad Request response with information regarding which parameters are invalid and why. After the \CoreAPIServer{} validates each request, it enqueues it in Redis for prioritization according to the priority rules. For example, authenticated users have higher priority. Once the measurement results are ready, \GlobalpingSystemName{} makes them available to the user or to \FrontEnd{} via HTTP polling, while logging every response and its outcome for audit purposes.

\subsubsection{Measurement Assignment Algorithm}

For each incoming measurement request, \GlobalpingSystemName{} filters probes based on the user's geographic constraints, and it filters out offline probes and those that have reached their maximum concurrent measurement limit. Finally, \GlobalpingSystemName{} ranks the probes using a weighted load score based on CPU and memory utilization and the current active measurement count and assigns measurements to the probes with the lowest load score until the requested number of probes is reached. If any probe becomes unresponsive during measurement task scheduling, \GlobalpingSystemName{} will move to the next-lowest-load-score probe in the appropriate region. 

\subsection{Measurement Execution}

For ping measurements, the probes of \GlobalpingSystemName{} measure the RTT of packets. RTT is the sum of the time it takes for packets to reach the destination and return to the source. \GlobalpingSystemName{} sends several packets and records their statistics, including packet loss rates and minimum, average, and maximum RTT values. Probes can also perform traceroute measurements by sending a set of packets with incremented TTLs. Each router (hop) along the packet's path decrements the TTL. Once each packet's TTL reaches zero, the router where it reaches zero sends a message back to the originating probe. The message contains the router's information, including its geographic location. Traceroute uses these messages to record the path of the packet. In addition, \GlobalpingSystemName{} can check a website's availability and response times by issuing a GET or HEAD request to an identified URL. \GlobalpingSystemName{}'s HTTP requests can also track additional metrics, such as time to DNS resolution and TCP connection establishment.

\subsection{Measurement Result Aggregation and Analysis}

Many probes may perform user measurements. Measurement results from probes exist as JSON objects. The probes send the measurement results, along with their unique identifiers, to \ProbeManagementModule{}, and \CoreAPIServer{} then sends them to Redis. A centric measurement repository combines them in Redis.

\GlobalpingSystemName{}'s \FrontEnd{} can visualize measurement results in real time. In \FrontEnd{}, a map shows the exact locations of the probes that execute the measurements, along with their packet latency, loss, and response times. \GlobalpingSystemName{} can also integrate with Slack and GitHub to display measurement results in their workspaces or in GitHub issue logs related to network performance.

\subsection{Scalability and Performance Optimization}
\GlobalpingSystemName{} uses load balancers that spread the incoming requests to multiple \CoreAPIServer{} instances, reducing the risk of bottleneck. \GlobalpingSystemName{} also monitors API performance and probe health using New Relic.

\GlobalpingSystemName{}'s API can support concurrent requests through queuing and asynchronous processing. \GlobalpingSystemName{} uses Redis for caching, reducing database load and read traffic. Probes can also automatically adjust the number of measurements they handle based on their host machine's CPU load.

\subsection{Security Considerations}

\GlobalpingSystemName{} probes are not general-purpose traffic generators to prevent them from being used like a botnet. Probes execute only predefined measurement commands, such as ICMP echo, traceroute, DNS lookups, and limited HTTP requests, and cannot send arbitrary packets, payloads, or high-bandwidth traffic. The \CoreAPIServer{} strictly bounds measurement parameters such as packet count, concurrency, and protocol types.

To prevent abuse, Globalping has per-user and per-IP rate limits, caps the number of probes that can participate in a single measurement, and limits the number of concurrent measurements per probe. By default, all users can make only a few requests per minute. These limits make DDoS or profit-driven traffic generation infeasible. Furthermore, \GlobalpingSystemName{} has up-to-date lists of domains and IP addresses known to host malware. These domains and private IP addresses are blocked to prevent malicious servers and unauthorized scanning of internal networks. 

The probes will also only issue an outgoing WebSocket connection to \ProbeManagementModule{}. There are no open ports on the probe, and it will never accept any incoming connections, limiting the risk of unauthorized access or attacks against the probe itself. The probes will communicate only with \ProbeManagementModule{} and perform the measurement. 

For resource management, volunteers can configure Docker to limit their CPU utilization; hence, no probe can ever overwhelm the host machine. Only one probe can run from each IP address to avoid resource misuse and coordinated attacks. \GlobalpingSystemName{} also logs access and usage patterns for probes.

\section{Evaluation}

\begin{table*}[ht]
\vspace{0.05in}
\centering
\small
\resizebox{\textwidth}{!}{
\begin{tabular}{| c | c | c | c | c | c | c |}
\hline 
& Official Integrations & Wider Target Audience & Community-Hosted Probes & Official Containerized Probes & Non-Expiring Free Tier & Open Source \\ [0.1ex] %
\hline
 RIPE Atlas & & & \checkmark & & \checkmark & \checkmark \\  
 PerfOps & & & & & \checkmark & \\
  Cedexis & & & & & & \\
   KeyCDN Tools & & & & & & \\
   \GlobalpingSystemName{} & \checkmark & \checkmark & \checkmark & \checkmark & \checkmark & \checkmark \\ [0.2ex]
\hline
\end{tabular}
}
\caption{\GlobalpingSystemName{}'s comparison to other tools in the industry}
\label{table:comparision}
\end{table*}

This section answers our research questions below:

\begin{itemize}
  \item \RQOne
  \item \RQTwo
  \item \RQThree
\end{itemize}

\subsection{\RQOne}
The tools closest to \GlobalpingSystemName{} we identified are RIPE Atlas, PerfOps, Cedexis, and KeyCDN Tools.

RIPE Atlas is a global Internet measurement network that collects data on Internet connectivity and reachability from thousands of probes worldwide \cite{b2}. Users can also earn credits in RIPE Atlas by hosting probes and spending them by requesting measurements from RIPE Atlas, similar to \GlobalpingSystemName{}. 

\GlobalpingSystemName{} and RIPE Atlas use a globally distributed network of community-hosted probes to perform ping, traceroute, and DNS resolution on target hosts from anywhere in the world \cite{b1, b2}. In addition, both have open-source components \cite{b1, b2}. RIPE Atlas also offers more sophisticated features, including custom measurements and anchored probes for continuous monitoring of critical points on the Internet \cite{b1}. RIPE Atlas has over 12,000 probes spread throughout the world, while \GlobalpingSystemName{} is still developing with +3000 probes \cite{b6}. The researchers published an academic paper in 2015 describing the technical details of RIPE Atlas \cite{b2}. This paper is the only scholarly work we have found in the literature on introducing a system that performs ND globally.

In terms of design, RIPE Atlas is more oriented towards research and academics \cite{b1}. Network engineers, researchers, and ISPs use it to make large-scale measurements \cite{b1}. On the other hand, \GlobalpingSystemName{} values simplicity and ease of use for all users, including researchers and non-technicals \cite{b1}. To the best of our knowledge, \GlobalpingSystemName{} is the first and only global ND platform to focus on democratizing ND through low-barrier access, officially using workflow integration tools, enabling real-time measurements instead of RIPE Atlas' near real-time, and officially using Docker for its probes, simplifying the setup process significantly compared to the hardware probes often associated with RIPE Atlas.

\GlobalpingSystemName{}'s primary use cases are benchmarking DevOps, monitoring CDN, and enabling teams to troubleshoot in real time \cite{b1}. One can run tests from a particular cloud vendor, such as AWS or Google Cloud, using \GlobalpingSystemName{} for specific regions to compare latency and connectivity and decide on the best vendor for one's requirements. Even running tests with \GlobalpingSystemName{} from certain ISPs can help debug ISP-level routing or performance issues affecting some users. Another use case is confirming that their CDN is properly routing traffic to the closest Point of Presence for users in various geographical areas and comparing performance across CDN vendors. In terms of reliability, \GlobalpingSystemName{} 's HTTP measurement functionality helps track uptime and response times for API endpoints across multiple worldwide locations. In the security realm, \GlobalpingSystemName{} can be used to verify the status and details of TLS certificates across various systems and confirm that they are correctly configured and have not expired.

\GlobalpingSystemName{} is intended for casual/commercial use and offers official integrations with platforms such as Slack and GitHub \cite{b1}. We are not aware of any official integrations for RIPE Atlas. However, users can still build custom tools with the RIPE Atlas API to send notifications or data to platforms such as Slack. Another flaw is that Bajpai et al. showed that the AS-based distribution of RIPE Atlas probes is heavily skewed, limiting the measurement capabilities of a specific origin AS \cite{b6}. We are not aware of a researcher who indicates a similar shortcoming to \GlobalpingSystemName{}.

In contrast to RIPE Atlas, PerfOps takes a more enterprise-oriented approach. PerfOps is a commercial network analytics tool that helps DNS and CDN providers troubleshoot their networks by collecting performance metrics and measuring resource usage \cite{b3}. DigiCert owns PerfOps and has more than 300 testing servers worldwide \cite{b3}. PerfOps offers companies comprehensive monitoring, benchmarking, and optimization of traffic routing on CDNs, DNS, and cloud services \cite{b3}. PerfOps also helps optimize traffic and perform load balancing to improve content delivery by analyzing and comparing various infrastructure providers \cite{b3}. On the other hand, \GlobalpingSystemName{} focuses on attracting all users with a simple interface and a low entry barrier. 

Another enterprise-oriented tool, Cedexis, is a SaaS-based Internet traffic management platform designed to offer automated, predictive, and cost-optimal routing of application, video, and web content \cite{b4}. Cedexis has been part of Citrix Systems since 2018. It enables companies to divert their traffic through any number of CDNs based on real-time performance metrics \cite{b4}. The core competencies of Cedexis include collecting real-time performance data and using that information to make intelligent routing decisions, ensuring that users' content comes from the fastest possible location \cite{b4}.

KeyCDN Tools is a set of online tools that help users analyze their website performance and identify connectivity issues \cite{b5}. It can perform operations similar to \GlobalpingSystemName{}, such as pinging and traceroute \cite{b5}. Its tools also help web developers and IT administrators who want to optimize web service delivery, troubleshoot CDN-related issues, and analyze HTTP performance \cite{b5}. Users don't have to host probes on KeyCDN Tools or set up Docker containers. Instead, KeyCDN's network contains all the probes. However, KeyCDN Tools only support traceroute from 14 locations, far fewer than \GlobalpingSystemName{} 's capabilities \cite{b7}.

Finally, we create a simple comparison illustration for \GlobalpingSystemName{} and its similar tools in Table \ref{table:comparision}. We did not include extensive tools, such as Datadog, in our comparison. Datadog can conduct API tests from different regions around the world. However, its primary purpose is application monitoring rather than ND.

Table \ref{table:comparision} uses our predefined and binary features for comparison. We define "Wider Target Audience" as whether the platform is designed to enable non-technical users to perform measurements. "Non-Expiring Free Tier" denotes whether the platform offers a non-expiring free usage tier. "Official Integrations" checks whether the platform's development team supports and maintains integrations like Slack. We also have "Official Containerized Probes" to check whether the platform supports running probes in containerized environments.

\subsection{\RQTwo}

\GlobalpingSystemName{} 's adoption is still growing. It has 3032 probes across 632 cities. The probes are spread across 114 countries and in 949 Autonomous System Numbers, indicating a high ISP reach. The researchers also used \GlobalpingSystemName{} and published their findings as a bachelor's thesis in the Faculty of Information Technology CTU in Prague \cite{b9}.

In addition, users initiate more than 300000 measurements each day through \GlobalpingSystemName{}'s API, \FrontEnd{}, and the CLI. We have not measured which channel users use the most to request measurements. However, we measured the average response time of \GlobalpingSystemName{}'s API, which was less than 14 milliseconds, indicating low latency. 

\subsection{\RQThree}
We conducted a System Usability Scale (SUS) study with 40 participants recruited via Profilic, an online platform that facilitates participant recruitment for online studies. We only required participants to speak English as their primary language. We did not set any other filters, such as technical background, networking experience, or demographics, to target non-technical users. Since \GlobalpingSystemName{} was new at the time of this study, we did not expect any participants to have prior exposure to it, so we did not control for it. Participants were randomly assigned to either the \GlobalpingSystemName{} (n=20) or RIPE Atlas (n=20) group. After the experiment ended, we found that participants' demographics on Prolific were randomly distributed across gender, ethnicity, nationality, student status, and employment status.

The SUS study was intentionally limited to a single, basic task, which is executing a ping measurement, to isolate perceived usability and initial accessibility, rather than to evaluate the system's full functional breadth or professional effectiveness. Our evaluation focuses on a ping-based task because it is the most fundamental and universally supported network diagnostic operation across global measurement platforms. This choice allows us to conduct a controlled, comparable usability assessment while minimizing the confounding complexity introduced by more advanced tasks.

To instruct the 20 participants who tested \GlobalpingSystemName{}, we used the prompt "Go to \GlobalpingSystemName{}'s homepage at https://globalping.io and try sending a ping from any location in Germany to example.com in 5 minutes. Then, answer the survey using the link." For RIPE Atlas users, we used the same prompt, except that we replaced the system name and its URL. The survey was conducted via Google Forms and included 10 SUS questions. We also added a question asking how long the ping operation took. We predicted that some users might not be able to complete the task, so that they might get confused by this question. To address this, we added the following information to the survey description: "You will still get compensated if you could not complete the task in 5 minutes. Please enter N/A to the second question (How many minutes did the task take you?) if you could not complete the task."

Table \ref{table:sus_results} indicates the SUS results we gathered from the 40 participants for both \GlobalpingSystemName{} and RIPE Atlas. We used a paired t-test and found a P-value of 0.00057. We also found the 95\% confidence interval: \GlobalpingSystemName{} will score between 13.13 and 33.37 points higher than RIPE ATLAS. In UX research, a Cohen's d above 0.8 is considered large, and ours is 1.64. \GlobalpingSystemName{} users took an average of 3.23 minutes to complete our ping task, and 3 users entered N/A, indicating they couldn't complete it. On the other hand, RIPE Atlas users took an average of 6.87 minutes to run a ping command, and 5 users couldn't do so, entering N/As. Although we asked users to spend only 5 minutes on the ping task, some users exceeded the time limit, with most spending 20 minutes.

\begin{table*}[h]
\vspace{0.05in}
\centering
\caption{SUS survey score results.}
\resizebox{\textwidth}{!}{
\begin{tabular}{|l|c|c|}
\hline
 Question & \GlobalpingSystemName{} Average Score &  RIPE Atlas Average Score\\ [0.5ex]
\hline
I think that I want to use this system frequently. & 3.25 & 2.5 \\
I found the system unnecessarily complex. & 2.1 & 3.65 \\
I thought the system was easy to use. & 4 & 2.2 \\
I think that I would need the support of a technical person to be able to use this system. & 2.5 & 3.55 \\
I found that the various functions in this system were well integrated. & 3.25 & 3.25 \\
I thought there was too much inconsistency in this system. & 2 & 2.9 \\
I would imagine that most people would learn to use this system quickly. & 3.75 & 2.95 \\
I found the system very cumbersome to use. & 2.9 & 3 \\
I felt very confident using the system. & 3.8 & 2.65 \\
I needed to learn a lot of things before I could get going with this system. & 2.3 & 3.5 \\
\hline
Total SUS Score: & 65.62 & 42.37 \\
\hline
\end{tabular}
}
\label{table:sus_results}
\end{table*}

\section{Discussion}

In this section, we will interpret the results, describe the limitations of our work, and discuss our plans for future work.

\subsection{Interpretation of Results}

We believe that \GlobalpingSystemName{} and RIPE Atlas share the most similar architecture due to their community-driven, open-source approach. Although RIPE Atlas is more established, \GlobalpingSystemName{} offers unique benefits and use cases that RIPE Atlas may not cover. \GlobalpingSystemName{} is a modern solution that focuses on user experience and casual/commercial use cases. \GlobalpingSystemName{} developers worked to make \FrontEnd{} intuitive to target every user base, including experts. This focus also makes it easier for non-technical users to run measurements quickly, making \GlobalpingSystemName{} more approachable to resource-constrained small businesses. They can lack technical experts, but the intuitive interface of \FrontEnd{} helps all users' learning curve. In addition, RIPE Atlas requires at least \$5000 to earn credits through donations, while \GlobalpingSystemName{} is only \$1 \cite{b10}.

Non-technical business users can also utilize \GlobalpingSystemName{}'s official integrations instead of building their custom applications to get data into their workspaces. Although RIPE Atlas and Cedexis may have community-developed integrations, official integrations are more stable, secure, and reliable because they are supported and updated by the platform's development team. Essentially, \GlobalpingSystemName{} democratizes ND access for all users from anywhere in the world. 

In addition to our comparative study, we used SUS for our user study. In the SUS scoring system, the 50th percentile is 68, and the 10th percentile is 51. \GlobalpingSystemName{} scored 65.62, just below the 50th percentile, while RIPE Atlas was 42.38, well below the 10th percentile. This result shows us that users were more comfortable with \GlobalpingSystemName{}. However, 65.62 is still a low SUS score for a user interface. \GlobalpingSystemName's goal is to give anyone, regardless of background, the ability to run network diagnostic commands. Thus, we deliberately did not use any screening or prequalification when selecting participants, and participants' primary language was English. The lack of user selectivity may have allowed those without computer knowledge to take our test. We aimed to enable non-technical users to take the test, and sending a ping command from \GlobalpingSystemName{} may have looked challenging to them. We consider it a success that we score just below the 50th percentile among non-technical users, compared to RIPE Atlas, which is much lower than the 10th percentile.

Although measuring user comfort with SUS was an essential part of our work, we also examined functionality. We measured the number of users who couldn't complete our ping task and the time it took the others to complete it. \GlobalpingSystemName{} participants completed the ping task in an average of 3.23 minutes, and only 3 of 20 users were unable to complete it. We consider this a very short timeframe for non-technical users to run a ping command from Germany to example.com. Compared to RIPE Atlas' 6.87 minutes on average, we believe \GlobalpingSystemName{} significantly expedites ND via ping and enables more users to run it successfully.

\GlobalpingSystemName{} exceeded RIPE Atlas in user experience but is still well behind in adoption and usage. RIPE Atlas has about 12 times as many probes as \GlobalpingSystemName{}. However, the barrier to entry for it is much higher than \GlobalpingSystemName{} as it does not even require the user to create an account to run network commands. The easy \FrontEnd{} and low barrier to entry will accelerate \GlobalpingSystemName's adoption and make it comparable with the RIPE Atlas community in a few years.

\subsection{Limitations}

We picked random users on Prolific with non-technical backgrounds, as we expect technical users to use network diagnostic tools more. Network commands may not have a significant use case for people outside the software profession. Therefore, selecting random participants did not fully demonstrate \GlobalpingSystemName's capabilities. We could have had better results with our SUS survey if we had chosen software engineers as participants.

After having more technical users, we could address another shortcoming. We only gave users a ping task, testing only a section of \GlobalpingSystemName{}. We could have asked the users to test more complex tasks, such as determining a packet's path or other commands. Adding more tasks could allow us to draw more granular insights on \GlobalpingSystemName{}.

Although our adoption numbers are promising, they are still in the early stages. This fact raises questions about sustainability and widespread adoption in the community. The size of the problem will decrease over time as the \GlobalpingSystemName{} community grows.

\subsection{Future Work}

We plan to conduct additional UX surveys, including the Single-Ease Question and the User Effort Score, across a broader range of tasks, such as traceroute, to gather user feedback and make \GlobalpingSystemName{} easier to use. We also plan to conduct additional quantitative studies to provide more robust evidence on the low latency and ease of use of \GlobalpingSystemName{}. 

In addition, Prolific allows study designers to select participants who are software engineers or who use technology for a certain amount of time in their work. We could categorize our participants by technical level and measure whether their comfort gap between \GlobalpingSystemName{} and the RIPE Atlas increases as users become more technical. We could even draw deeper insights from having different software engineering specialties, such as network administrators, trying \GlobalpingSystemName{}. However, we did not see this level of granularity on Prolific, so we would need to improve our user-study capabilities.

Although we are satisfied with our SUS score and have not tested the \GlobalpingSystemName{} \FrontEnd{} with software engineers, we could implement guided tutorials, or onboarding flows to help new users quickly understand and utilize Globalping's features. We could also have participants test \GlobalpingSystemName{} both with and without tutorials to measure their impact.

\section{Conclusion}

\GlobalpingSystemName{} enables technical and non-technical users to apply scalable, community-driven, open-source, and real-time network diagnostics via a globally distributed network of probes, allowing them to perform a wide range of network measurements, such as ping, traceroute, and DNS lookups. These features, in combination with official integrations such as Slack and GitHub, help both casual users and team members perform real-time network troubleshooting. Compared with similar tools, such as RIPE Atlas, \GlobalpingSystemName{} targets all users, including experts and non-technical users, while maintaining robust real-time monitoring and diagnostics capabilities. We believe \GlobalpingSystemName{} is a valuable tool for performance measurement, offering powerful, user-friendly solutions worldwide.

\section*{Acknowledgment}

The authors thank Dmitriy Akulov for building and supporting this project.

\end{document}